\documentclass[notitlepage,aps,pra,twocolumn,epsfigure,superscriptaddress,longbibliography,nofootinbib]{revtex4-1}

\usepackage{latexsym}
\usepackage[latin9]{inputenc}
\usepackage{mathtools}
\usepackage{amsbsy}
\usepackage{amstext}
 \usepackage[colorlinks=true,linkcolor=NavyBlue,urlcolor=blue,citecolor=blue,pdfusetitle]{hyperref}
\usepackage[dvipsnames]{xcolor}
\usepackage{natbib}
\usepackage{amsmath}
\usepackage{graphicx}
\usepackage{amssymb}
\usepackage{txfonts,color}

\makeatletter

\@ifundefined{textcolor}{}
{%
 \definecolor{BLACK}{gray}{0}
 \definecolor{WHITE}{gray}{1}
 \definecolor{RED}{rgb}{1,0,0}
 \definecolor{GREEN}{rgb}{0,1,0}
 \definecolor{BLUE}{rgb}{0,0,1}
 \definecolor{CYAN}{cmyk}{1,0,0,0}
 \definecolor{MAGENTA}{cmyk}{0,1,0,0}
 \definecolor{YELLOW}{cmyk}{0,0,1,0}
}

\newcommand\ket[1]{\left|#1\right\rangle}
\newcommand\bra[1]{\left\langle #1 \right|}

\makeatother

\begin{document}
\title{Open Quantum Rotors: Connecting Correlations and Physical Currents}
\author{Ricardo Puebla}\email{These authors contributed equally to this work}
\affiliation{Instituto de F{\'i}sica Fundamental, IFF-CSIC, Calle Serrano 113b, 28006 Madrid, Spain}
\affiliation{Centre for Theoretical Atomic, Molecular, and Optical Physics, School of Mathematics and Physics, Queen's University, Belfast BT7 1NN, United Kingdom}
\author{Alberto Imparato}\email{These authors contributed equally to this work}
\affiliation{Department of Physics and Astronomy, University of Aarhus, Ny Munkegade, Building 1520, DK-8000 Aarhus C, Denmark}
\author{Alessio Belenchia}
\affiliation{Institut f\"{u}r Theoretische Physik, Eberhard-Karls-Universit\"{a}t T\"{u}bingen, 72076 T\"{u}bingen, Germany}
\affiliation{Centre for Theoretical Atomic, Molecular, and Optical Physics, School of Mathematics and Physics, Queen's University, Belfast BT7 1NN, United Kingdom}
\author{Mauro Paternostro}
\affiliation{Centre for Theoretical Atomic, Molecular, and Optical Physics, School of Mathematics and Physics, Queen's University, Belfast BT7 1NN, United Kingdom}


\begin{abstract}
We consider a finite one-dimensional chain of quantum rotors interacting with a set of thermal baths at different temperatures. When the interaction between the rotors is made chiral, such a system behaves as an autonomous thermal motor, converting heat currents into non-vanishing rotational ones. Such a dynamical response is strongly pronounced in the range of the Hamiltonian parameters for which the ground state of the system in the thermodynamic limit exhibits a quantum phase transition. Such working points are associated with 
large quantum coherence and multipartite quantum correlations within the state of the system.  
 This suggests that the optimal operating regime of such quantum autonomous motor is one of maximal quantumness.

\end{abstract}

\date{\today}

\maketitle


\section{Introduction}


There is a growing, cross-disciplinary interest in the understanding of the way quantum features affect the laws of thermodynamics~\cite{Batalhao:14,Brandao:15,Alhambra:16,Brunelli:18,Landi:20} and explore the limits to thermal machines operating at the nano-scale~\cite{Kosloff:84,Linden:10,Abah:12,Rosnagel:14,Rosnagel:16,Peterson:19,Lindenfels:19}. 
While, so far, the focus of such investigations has been primarily put on simple quantum systems involving only a few degrees of freedom,  the assessment of the thermodynamic performance of quantum many-body systems as working media of potential quantum motors have recently started to receive attention~\cite{Dorner:12,Watanabe:20,Hartmann:20,Fei:20,Fogarty:20,Revathy:20}. 

Autonomous thermal motors are of particular interests for the thermodynamics of both classical \cite{Reimann07, Dotsenko13, Argun17, Mancois18,Fogedby17,Imparato21} and quantum processes \cite{Mari15,Roulet17,Seah18,Fogedby18,Roulet2018,Hovhannisyan:19,Drewsen19}. Such devices are able to convert thermal currents into motion, and thus possibly work.
Their most salient feature is that they can operate without the intervention of an external agent that changes their Hamiltonian, making their design ideal for application purposes.
Autonomous quantum refrigerators have a similar task, cooling down a reservoir at the expenses of heath currents \cite{Brunner16,Hewgill20,Hewgill21}. 

Recent work has shown that collective phenomena such as synchronization and classical phase transitions can enhance the dynamic and thermodynamic performances in systems of interacting molecular motors~\cite{Golubeva2012a,Golubeva2013,Golubeva2014},
 of interacting  work-to-work transducers~\cite{Imparato15,Herpich18,Herpich18a},  in a 2D system of classical rotors driven out of equilibrium by a temperature gradient~\cite{Sune19a}, or in an out-of-equilibrium Frenkel-Kontorova model undergoing a commensurate-incommensurate phase transition~\cite{Imparato21}. These are a fascinating phenomena arising from the collective behavior in a many-body system~\cite{Huang}, which divide the phases of matter characterized by different properties depending on the external conditions. This phenomenon also applies to quantum systems, where quantum fluctuations -- rather than thermal ones -- can trigger quantum phase transitions (QPTs)~\cite{Sachdev}. At a quantum critical point, the ground state of the system develops singular behavior, typically accompanied by the closing of the energy gap~\cite{Sachdev} with the first excited state and diverging quantum correlations~\cite{Osborne:02,Vidal:03,DeChiara:18}, among other features. 
 
 The study of autonomous thermal motors and refrigerators based on quantum many-body effects could thus potentially allow for the identification of possible performance enhancements stemming from collective quantum phenomena such as a QPT. In this paper, we investigate the thermodynamics of an autonomous system in proximity of a QPT. We consider a finite-size one-dimensional chiral clock model (CCM) consisting of interacting quantum rotors. In the thermodynamic limit of infinitely many constituents, this model exhibits a well-characterized QPT ~\cite{Fendley:12,Ortiz:12,Whitsitt:18,Samajdar:18}. A dimer of quantum rotors with chiral interaction has been shown to give rise to a rotational current, when connected to two baths at different temperatures, as a result of the lack of thermal equilibrium and owing to the broken rotational symmetry~\cite{Hovhannisyan:19}. 
In the multi-component system considered here, we find that such a dynamical response is maximal for values of the Hamiltonian parameters that result in a QPT in the thermodynamic limit. Although the rotational current turns out to  always be finite, such a phenomenon is reminiscent of the diverging response to a change in an external thermodynamic force in systems at equilibrium in proximity of a phase transition, a phenomenon whose onset we are able to witness despite the finiteness of the system that we address. Furthermore we elucidate the relation between quantum correlations and thermodynamic currents in the considered CCM. While the unveiled phenomenology does not imply necessarily a causal link between the emergence of mechanical currents and the onset of many-body criticality, the interplay between these effects is suggestive of a strong role played by collective phenomena on the performance of heat-to-mechanical current conversion in such autonomous device. 


The remainder of this paper is organized as follows. In Sec.~\ref{iso} we introduce the basics of the CCM. In Sec.~\ref{open}, we consider the interaction of the rotors with independent thermal baths with staggered temperatures. We characterize the non-equilibrium  steady-state (NESS) of the model by looking at the tunneling and thermal currents. In Sec.~\ref{curr}, we connect the particle currents at the steady-state with the correlations established within the clock model. Finally, in Sec.~\ref{conc} we summarize the main findings reported in the article.

\section{Chiral Clock Model: Quantum phase transition in the isolated system}\label{iso}

Let us start by considering the $\mathbb{Z}_{N_s}$ CCM for $M$ quantum rotors~\cite{Fendley:12,Ortiz:12,Whitsitt:18,Samajdar:18,Hovhannisyan:19}, i.e. $M$ quantum systems with $N_s$ discrete energy levels. Each individual rotor can be seen as a spin-$(N_s - 1)/2$ or as particles occupying the $N_s$ vertexes of a regular polygon. Let $\{|k\rangle_i\}$ denote the orthogonal basis of the Hilbert space of the $i^\text{th}$ rotor, with $k = 0, . . .N_s - 1$ corresponding to the $N_s$ directions along which the angular momentum can point (or vertices of the polygon), with $\ket{k+N_s}_i=|k\rangle_i$. The CCM is then described by the Hamiltonian
\begin{equation}\label{eq:Hccm}
H_{\rm ccm}=-f\sum_{j=1}^M(\sigma_j +\sigma_j^\dagger )-(1-f)\sum_{j=1}^{M}\left(\mu_{j}\mu_{j+1}^{\dagger}e^{i\varphi_j}+h.c.\right),
\end{equation}
where $f$ is the control parameter that accounts for the relative weight between the free and interaction terms,  and $\varphi_j$ the so-called chiral phases.
We assume periodic boundary conditions, so that $\mu_{M+1}=\mu_1$.
Here the operators $\mu$ and $\sigma$, in the vertexes basis $\{|j_1,\cdots,j_M\rangle\}$, are defined as
\begin{align}
\mu=\begin{bmatrix} 
1 & 0 & 0 & \cdots & 0\\ 
0 & \omega & 0 & \cdots & 0\\ 
0 & 0 & \omega^2 & \cdots & 0\\
0 & 0 & 0 & \ddots & 0\\
0 & 0 & 0 & 0 & \omega^{N_s-1}
\end{bmatrix}, 
\quad \sigma=
\begin{bmatrix} 
0 & 1 & 0 & 0 & \cdots & 0\\ 
0 & 0 & 1 & 0 & \cdots & 0\\
0 & 0 & 0 & 1 & \cdots & 0\\
0 & 0 & 0 & 0 & \ddots & 0\\
0 & 0 & 0 & 0 & \cdots & 1\\
1 & 0 & 0 & 0 & \cdots & 0
\end{bmatrix}
\end{align}
with $\omega=e^{i2\pi/N_s}$. The first term of the Hamiltonian encodes the dynamics of the individual rotors and gives rise to tunneling currents between their internal levels (cf. Fig.~\ref{fig1}). For  a particle at the vertices of a regular polygon, the tunneling currents can be visualised as describing the hopping of the single system between such vertices induced  by the rotor internal Hamiltonian. The second term in the Hamiltonian encodes the interaction between nearest neighbors. 

The model possesses a global $\mathbb{Z}_{N_s}$ symmetry and, classically, presents two phase transitions in 2D \cite{Lapilli06}. The interaction potential breaks a specific rotational symmetry when $\varphi_j\neq k\pi/N_s$ ($k\in\mathbb{Z}$), as discussed in Refs.~\cite{Hovhannisyan:19,Hewgill21}. This is a necessary condition for the emergence of the rotational (particle) currents, as we will also see in the following  (cf. Sec.~\ref{nessgkls}). In this context, the order parameter of the model is the total magnetization $m=\sum_j(\mu_{j}+\mu_{j}^\dagger)/M$.
From now on, we will focus on the minimal configuration allowing for non-zero currents, namely the case of $N_s=3$. It should be noted that, our model is similar to the one investigated in Ref.~\cite{Samajdar:18} where the role of $\sigma$ and $\mu$ was interchanged. In Refs.~\cite{Whitsitt:18,Samajdar:18}, the structure of the phase diagram of the CCM with $N_s=3$ and homogeneous chiral phase $\varphi_j=\varphi$ was investigated in detail, showing that for small values of $\varphi$ there is a direct transition from the ordered ($f\ll 1/2$) to a disordered phase ($f\gg 1/2$). For large chirality ($\varphi>\pi/6$), the two phases are separated by an incommensurate  phase.
In Appendix~\ref{app:a} we provide a brief summary of the symmetry-breaking QPT taking place in the ground state of $H_{\rm ccm}$, while we refer to Refs.~\cite{Whitsitt:18,Samajdar:18} for a thorough inspection of the model's critical features.

\section{Open System Dynamics: Correlations vs Currents}\label{open}

\begin{figure}
\centering
\includegraphics[width=1\linewidth,angle=-0]{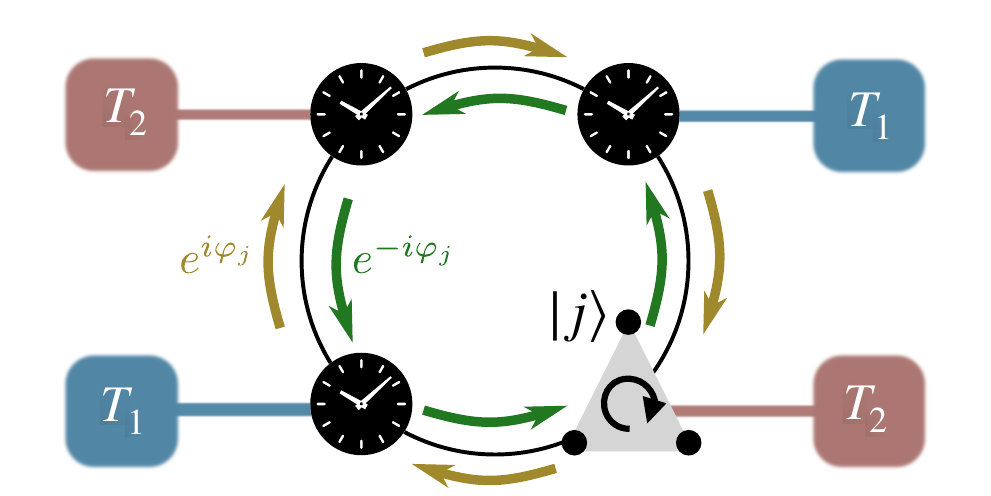}
\caption{\small{Schematic representation of the CCM with periodic boundary conditions, interacting with thermal reservoirs with staggered temperatures $T_1$ and $T_2$, and chiral hopping interactions governed by the phase $\varphi_j$.}}
\label{fig1}
\end{figure}

We are interested in exploring the physics of the CCM when interacting with thermal baths. In particular, we consider the case in which each rotor is in contact with an independent thermal reservoir and partition our system in two sub-lattices consisting of even (e) and odd (o) rotors, respectively. The inverse temperature of the two sub-lattices is set to be $\beta_e$ and $\beta_o$, respectively, and we will assume $\beta_e\neq \beta_o$, in general, thus realizing a staggered-temperature configuration (cf. Fig.~\ref{fig1} for a schematic illustration). As it will be shown later on in this Section, the temperature difference gives rise to thermally driven mechanical currents in the system that are sustained asymptotically in time. The system thus evolves towards a non-equilibrium steady-state (NESS), whose properties we now aim at characterizing.

We describe the open system dynamics via the local Gorini-Kossakowski-Sudarshan-Lindblad (GKLS) master equation 
\begin{equation}
    \dot\rho=-i[H_{\rm ccm},\rho]+\sum_{m=1}^M\mathcal{D}_m(\rho),
    \label{mas:eq}
\end{equation}
with local dissipators
\begin{align}
\mathcal{D}_m[\bullet]=\sum_{j,j'} W_{j,j'}^{(m)} \left[L_{j,j'} \bullet L_{j,j'}^\dagger-\frac{1}{2}\left\{L_{j,j'}^\dagger L_{j,j'},\bullet\right\} \right].
\end{align}
defined in terms of the jump operators $L_{j,j'}=\ket{j}\bra{j'}$ with  $\ket{j}=\ket{j_1,\ldots,j_M}$ with $j_k = 0,\cdots, N_s-1$. The transition rates $W_{j,j'}$ from $\ket{j'}$ to $\ket{j}$ 
fulfill the local detailed balance
\begin{align}
{W_{j,j'}^{(m)}}/{W_{j',j}^{(m)}}=e^{\beta_m (E_{j'}-E_{j})},
  \end{align}
where $E_{j}=\bra{j}H_{\mathrm {ccm}}\ket{j}$. For a generic bosonic bath, we have $W^{(m)}_{j,j'}=\gamma_m(E_{j'}-E_j)$ with~\cite{Breuer}
\begin{align}\label{eq:gammaw} 
  \gamma_m(\omega)=\frac{g|\omega|}{1-e^{-\beta_m |\omega|}}\zeta(\omega)\quad\text{and}\quad\zeta(\omega)=\left\{\begin{matrix}  e^{\beta_m \omega} &\omega\leq 0, \\ 1 & \omega>0 , \end{matrix}\right.
  \end{align}
and where $g$ is a microscopic rate.
Before proceeding further, a note is in order. As it is well known, local master equations can be problematic from a thermodynamic point of view~\cite{Levy14,Stockburger17}. However, it should be noted that this conclusion has been  recently challenged by a stream of works~\cite{Barra15,Strasberg17,DeChiara1811,Hewgill21} pointing towards a reconciliation of local master equation and thermodynamics. In particular, it has been shown that the local master equation is not, in general, at odds with the second law of thermodynamics as far as the proper  expression for the  heat currents is considered. In the specific case under study, we can split the Hamiltonian in its diagonal and non-diagonal part in the $\{|j\rangle\}$ basis as $H_{\rm ccm}= H_D + H_{ND}$, which allows us to introduce the individual energy currents
\begin{equation}
\label{diagandnondiag:Q}\
\begin{aligned}
    \dot{Q}_{D,m}&={\rm tr}(\rho\mathcal{D}_m^*[H_D]),\\
    \dot{Q}_{ND,m}&={\rm tr}(\rho\mathcal{D}_m^*[H_{ND}]),
\end{aligned}
\end{equation}
where $\mathcal{D}_m^*$ is the dual of $\mathcal{D}_m$. It is useful to remark that the {\it standard} definition of heat flux when dealing with a local master equation would read $\dot{Q}_m={\rm tr}(\rho\mathcal{D}_m^*[H_{\rm ccm}])=\dot{Q}_{D,m}+\dot{Q}_{ND,m}$. Unfortunately, using $\dot{Q}_m$ leads in general to violations of the second law of thermodynamics (cf. Ref.~\cite{Levy14} for an example). However, it is the weighted sum of $\dot{Q}_{D,m}$'s that enters the second law of thermodynamics and gives a positive entropy production rate $\dot{\Sigma}=dS/dt-\sum_m\beta_m\dot{Q}_{D,m}\geq 0$, consistently with the second law~\cite{Hewgill21}, and one should really focus on the individual currents. On the other hand, the $\dot{Q}_{ND,m}$ can be associated to a work rate within a microscopic collisional model framework~\cite{DeChiara1811}. For further details on this contruction we refer the  interested reader to~\cite{Hewgill21}.

\subsection{NESS of the GKLS Master Equation and Particle Currents}\label{nessgkls}
From the numerical diagonalization of the Lioville super-operator on the right-hand-side of the GKLS master equation, we obtain the unique steady-state $\rho_{\rm SS}$ of the CCM interacting with independent thermal baths. Such state is in general a NESS, however its nature is determined by the choice of parameters of the model. Note that, although the ground state of $H_{\rm ccm}$ displays a QPT, such abrupt transition is blurred in this open quantum system setting. 
In order to quantify the non-equilibrium nature of the steady-state we turn to look  at quantum particle currents in the system.

The definition of quantum particle currents in general is a non trivial task. A formal characterisation has been carried out in \cite{Hovhannisyan:19} where the authors also investigate a CCM with $M=2$ rotors. 
For a classical particle hopping on a graph, one can readily define the probability current between any two vertices on the graph which reads
\begin{equation}
    J_{j \rightarrow j'} = W_{j'j}p_j -W_{jj'}p_{j'},
\label{J:class}
\end{equation}
 where $p_j$ is the instantaneous probability of finding the particle at vertex $j$, and $W_{jj'}$ is the transition rate from $j'$ to $j$.
In Ref.~\cite{Hovhannisyan:19}, the quantum analogous of this classical current was defined as the sum of the tunnelling and thermal current operators, namely, 
\begin{align}\label{eq:Jtun}
&J^{\rm tun}_{j \rightarrow j'} 
=  i(x_j H_{\rm ccm} x_{j'} - x_{j'} H_{\rm ccm} x_j),\\ \label{eq:Jth}
&J^{\rm th}_{j \rightarrow j'}
=\frac{1}{2}\sum_\lambda \gamma_\lambda \left[\left\{x_j,L^\dag_\lambda x_{j'}L_\lambda\right\}-\left\{x_j',L^\dag_\lambda x_{j}L_\lambda\right\}\right],
\end{align}
where $x_j = \ket{j}\bra{j}$ is the projector onto a generic state $\ket{j}$, and the sum runs over all the possible transitions $\lambda$ between pair of states of the system. 
Note that the thermal current reduces to the classical probability current (\ref{J:class}) in the classical limit.
Furthermore, in order to simplify the analysis of the dynamics, in the following we will only allow jumps between states where only one
spin is rotated, that is $\ket{j}\equiv\ket{j_1,\ldots j_l\ldots ,j_M} \to \ket{j'}\equiv\ket{j_1,\ldots j_l\pm 1\ldots,j_M}$.


\begin{figure}
\centering
\includegraphics[width=1\linewidth,angle=-0]{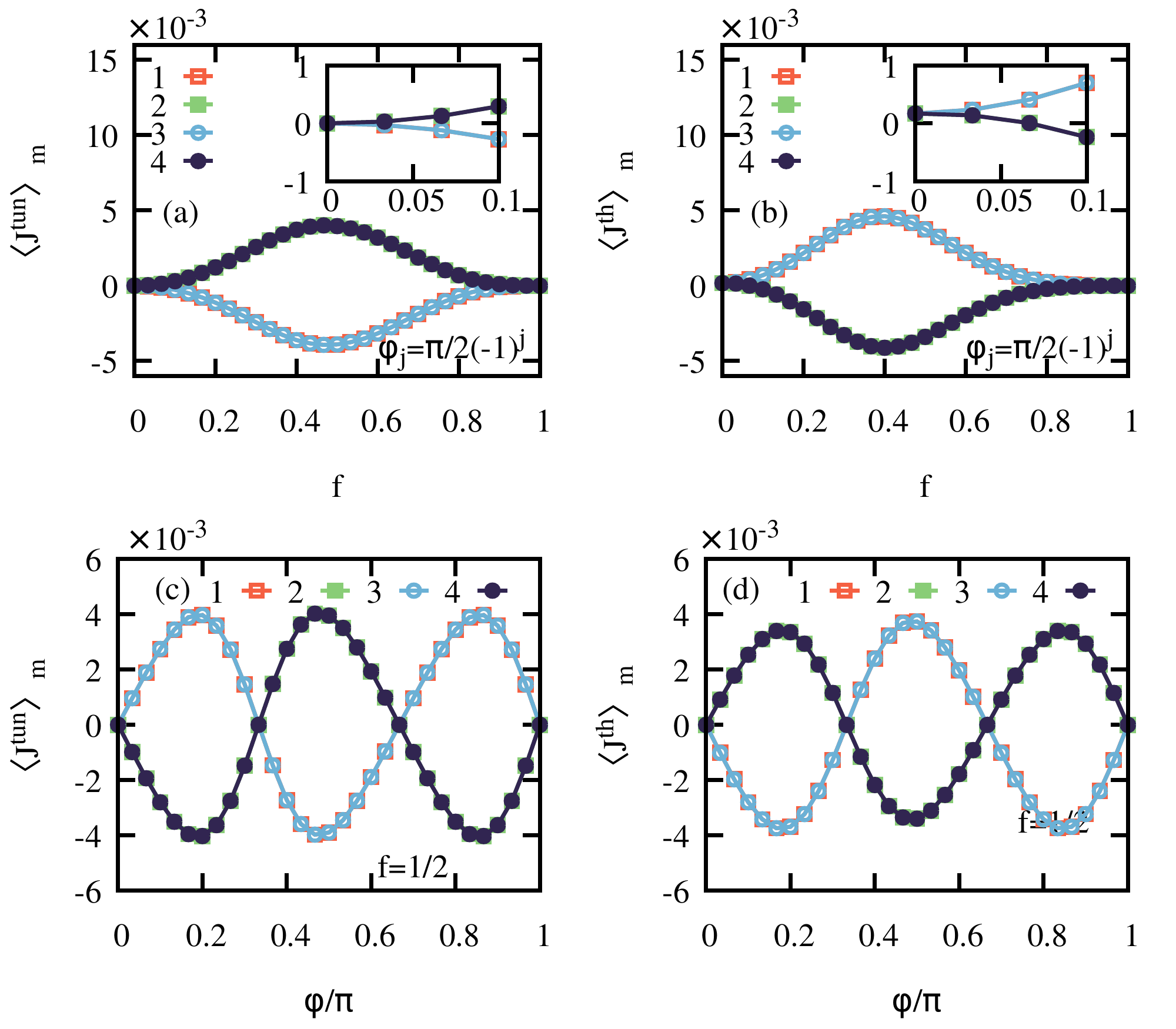}
\caption{\small{Illustration of the non-trivial behavior of the NESS tunneling and thermal currents $\langle J_{0\to1}^{\rm tun}\rangle_m$ and $\langle J_{0\to1}^{\rm th}\rangle_m$ for each rotor in a chain with $M=4$ and $N_s=3$. Panels (a) and (b) [(c) and (d)] illustrate such currents as functions of $f$ [$\varphi$ with $\varphi_j=(-1)^j\varphi$] for $\beta_e=1$ and $\beta_o=1.1$ with $g=0.2$. Panels (a) and (b) have been obtained taking fixed staggered chiral phases $\varphi_j=(-1)^j\pi/2$, while the insets show the behavior of currents in the region close to $f=0$. Note that 
$\langle J_{0\to1}^{\rm tun}\rangle= 0$, while $\langle J_{0\to1}^{\rm th}\rangle\neq 0$ for $f=0$. Panels  (c) and (d) have been obtained taking $f=1/2$, and $\varphi_j=(-1)^j\varphi$, here the currents display a $2\pi/N_s$ periodicity in $\varphi$.  See main text for further details and Fig.~\ref{fig3} for the total currents.}}
\label{fig2}
\end{figure}

\begin{figure}
\centering
\includegraphics[width=1\linewidth,angle=-0]{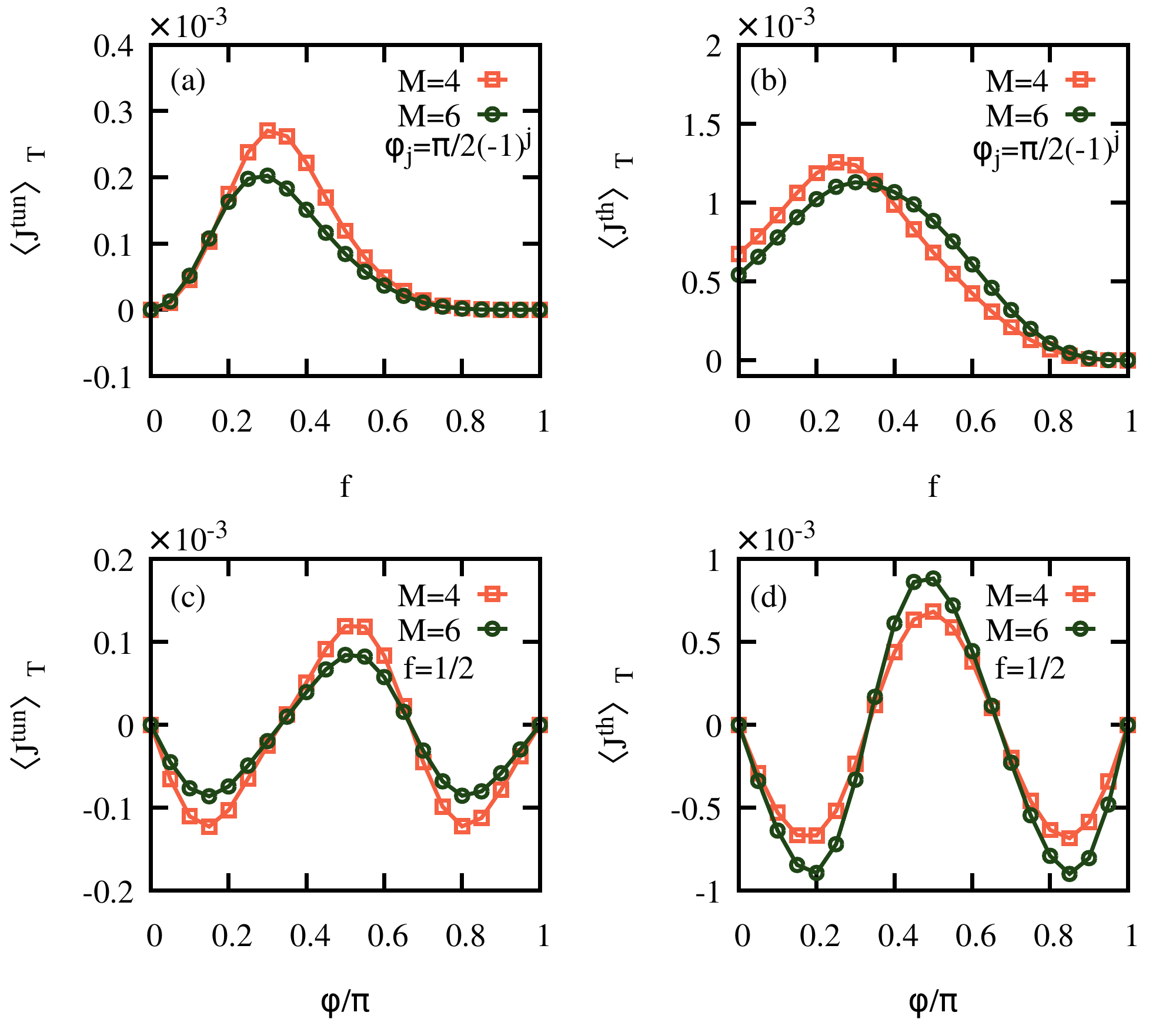}
\caption{\small{Total NESS currents $\langle J^{\rm tun}\rangle_T$ and $\langle J^{\rm th}\rangle_T$ for $M=4$ and $M=6$ rotors and $\beta_e=1$, $\beta_o=1.1$ with $g=0.2$.  Panels (a) and (b) are for $\varphi_j=(-1)^j\pi/2$, while panels (c) and (d) are for $f=1/2$.}}
\label{fig3}
\end{figure}

Before proceeding further, we shall discuss the general properties of these NESS currents. 
In order to simplify the notation, in the following we will omit the subscript $j\to j'$ in the steady state currents $\langle J^{\rm tun,th}_{j\to j'}\rangle$, as in the NESS the rotational currents of the $m^\text{th}$ rotor are independent of the specific initial and final position considered. 
For $\varphi=k\pi/N_s$ ($k\in \mathbb{Z}$) the currents vanish, $\langle J^{\rm tun}\rangle_m=\langle J^{\rm th}\rangle_m=0$, for each individual rotor $m=1,\ldots, M$, $\forall f$ and regardless of the temperature difference among sub-lattices $\Delta T=1/\beta_o-1/\beta_e$.

In addition, $\langle J^{\rm tun}\rangle_m\neq 0$ if $\varphi\neq k\pi/N_s$ and $f\neq 0,1$, and $\langle J^{\rm tun}\rangle_T=\sum_{m=1}^M \langle J^{\rm tun}\rangle_m=0$ when $\Delta T=0$. In a similar fashion, the thermal current fulfills $\langle J^{\rm th}\rangle_T=\sum_m \langle J^{\rm th}\rangle_m\neq 0$ for $\varphi\neq k\pi/N_s$ and $f\neq 1$, while $\langle J^{\rm th}\rangle_T=0$ for $\Delta T=0$. Fig.~\ref{fig2} show the individual steady-state currents as a function of $f$ and $\varphi$  (with $\varphi_j=(-1)^j \varphi$) for each of the individual rotors in a CCM with $M=4$, $N_s=3$, $\beta_e=1$, $\beta_o=1.1$ and $g=0.2$, which already reveal a non-trivial behavior. A similar behavior is found for different parameter combinations.
Inspection of Fig.~\ref{fig2} suggests that, as a function of $f$, the maximum of $\vert \langle J^{\rm tun}\rangle_m\vert$ is reached at $f\simeq 0.45$, which is very close to the value $f\simeq 0.46$ at which a QPT occurs in the ground state of the CCM at thermodynamic limit [cf. Appendix \ref{app:a}]. However, the thermal current $\langle J^{\rm th}\rangle_m$ is maximized for a slightly smaller value of $f$.
Also, the insets in Fig.~\ref{fig2} (a) and (b) show that, in the {\it classical} limit $f=0$, the tunnelling current is vanishing in both sub-lattices, while the rotors exhibit the same non-zero thermal rotational frequency.
Although not explicitly shown, the mean square value of the thermal current 
has a maximum at $f\approx 1/2$ while the analogous quantity for the thermal current gets the value of 
$\approx 2f^2/3$ independently of $\varphi$. This value suggests that all clock states are equally populated at the NESS. However, as shown in Sec.~\ref{curr}, such state is not a maximally mixed one as it 
brings about coherence and non-trivial correlations among the individual rotors. 

Fig.~\ref{fig3} also illustrates the total currents $\langle J^{\rm tun, th}\rangle_T$ for the same parameters as Fig.~\ref{fig2} for $M=4$ and $6$ rotors and $\Delta T\neq 0$. 
Note that for fixed $M$ the total thermal current is larger than the tunnelling one. Furthermore, for the two sizes here considered, $\langle J^{\rm th}\rangle_T$ is almost constant for increasing number of rotors, while $\langle J^{\rm tun}\rangle_T$ decreases its value suggesting that for large $M$ the total tunneling current will be negligible with respect to the thermal one. Hence, in the thermodynamic limit, one should expect  $\langle J^{\rm th}\rangle_T+\langle J^{\rm tun}\rangle_T\approx \langle J^{\rm th}\rangle_T$. 

\begin{figure}
\centering
\includegraphics[width=1\linewidth,angle=-0]{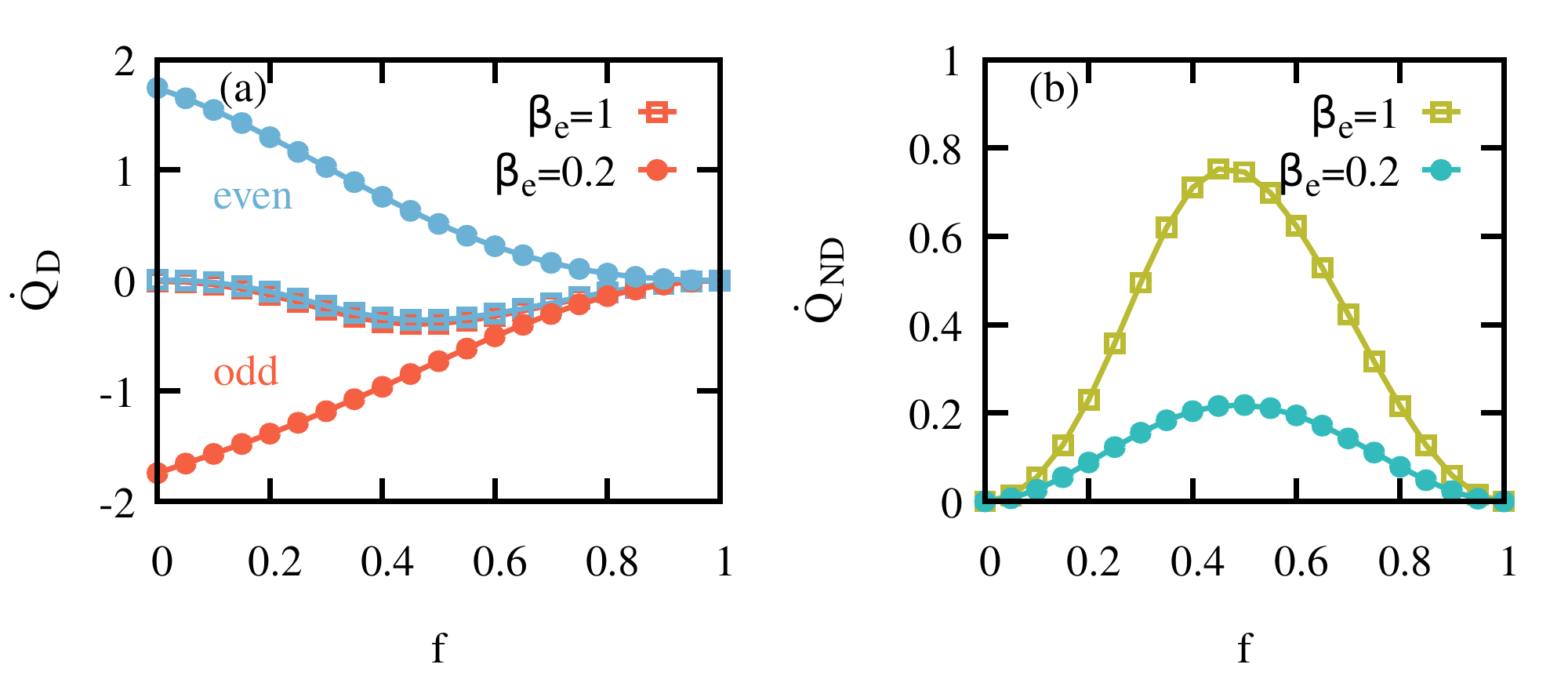}
\caption{\small{Diagonal (a) and non-diagonal (b) heat currents, $\dot{Q}_{{\rm D},m}$ and $\dot{Q}_{\rm ND}=\sum_m \dot{Q}_{{\rm ND},m}$, as defined by Eqs.~(\ref{diagandnondiag:Q}), respectively, as a function of $f$ for the NESS of the CCM with $M=4$ rotors and staggered chiral phases $\varphi_j=(-1)^j\pi/2$ and temperatures. We have taken $\beta_o=1.1$, $g=0.2$, and $\beta_e=0.2$ and $1$, as specified in the legend of the figure. In panel (a) we show the heat currents for each sub-lattice, namely, even and odd rotors. For $\beta_e=1$, $\dot{Q}_{\rm D}<0$ for both sub-lattices, which implies a large non-diagonal heat current. Note that $\sum_m\left(\dot{Q}_{{\rm D},m}+\dot{Q}_{{\rm ND},m}\right)=0$. }}
\label{fig4a}
\end{figure}

We now turn our attention to the steady state heat currents, as given by Eqs.~(\ref{diagandnondiag:Q}). Here the heat currents are positive when flowing from the bath(s) to the system.  The results, for two different sets of system parameters, are shown in Fig.~\ref{fig4a}. As previously done, we have chosen the even sub-lattice to be in contact with the hot bath. We observe that, for a small temperature  gradient, the diagonal heat currents are both negative. This can be understood as follows. First, the first law -- written in the form $\sum_m\left(\dot{Q}_{{\rm D},m}+\dot{Q}_{{\rm ND},m}\right)=0$ -- is valid. Second, we recall that the non-diagonal heat current $\dot Q_{\rm ND}$ corresponds, within the framework of the collisional model, to the work
done or produced when switching on and off the interaction
of the system with the colliding particles making up the environment~\cite{DeChiara1811,Hewgill21}. Thus, the situation in Fig.~\ref{fig4a} where $\dot{Q}_{{\rm D},m}<0$ for all rotors is compensated by a large and positive $\sum_m \dot{Q}_{{\rm ND},m}$, corresponding to a net amount of work done on the system that is then dissipated in both the cold and hot baths.
One can understand this result also noticing that, when $f>0$ the Hamiltonian in Eq.~(\ref{eq:Hccm}) is not diagonal in the basis $\ket j$. Thus, Eq.~(\ref{mas:eq}) will introduce coherence in the steady state, resulting in a non-zero non-diagonal heat current, as given by the second line of Eq.~(\ref{diagandnondiag:Q}).

For a larger temperature gradient, and one of the two temperatures relatively high,  the heath currents exhibits a more classical behaviour with a net diagonal current from the hot to the cold baths and a reduced non--diagonal heat current.

In the ground state of the system \eqref{eq:Hccm} the thermally driven current  $\langle J^{\rm th}\rangle_m$ vanishes for any $f$: At $T=0$ there is no heat current to sustain the rotational motion.
However, the tunnelling current  $\langle J^{\rm tun}\rangle_m$ may in principle be non-vanishing in the ground state: Eq.~\eqref{eq:Jtun} is indeed the discrete counterpart of the Schr\"odinger
probability current, as discussed in \cite{Hovhannisyan:19}. We find nevertheless that also  $\langle J^{\rm tun}\rangle_m$ vanishes in the ground state of \eqref{eq:Hccm} for any $f$.
In Appendix \ref{app:c} we consider the rotated model of \eqref{eq:Hccm} with $\sigma\rightarrow \mu$ and $\mu\rightarrow \sigma$, and interestingly find that the  $\langle J^{\rm tun}\rangle_m$ shown a critical-like behaviour in the ground state, being non-zero for $f\lesssim f_c$.

\section{Connecting currents to collective information theoretic quantities}\label{curr}

The connection between the location, in parameter space, of the quantum critical point of the CCM and that of the optimal particle currents is suggestive of a potential role of collective quantum phenomena in the establishment of the non-equilibrium features of the system. In this Section we explore such suggestion further by making use of a toolbox of information theoretic figures of merit that have been used, in the past, to explore the interplay between quantum critical phenomena and non-classicality~\cite{Campbell:13, Campbell:11,DeChiara:18}. In doing so, we unveil the intrinsically collective nature of the features that have been highlighted in our analysis so far.

Quantitatively, we will consider the von Neumann entropy of a subsystem $A$ of a compound $A\cup B$, which is defined as
\begin{align}\label{eq:Sa}
S_{A}=-{\rm Tr}[\rho_A\log \rho_A],
  \end{align}
where $\rho_A={\rm Tr}_{B}[\rho]$ denotes the partial trace over $B$. Another relevant measure is the negativity $N_A$~\cite{Vidal:02}, which is able to quantify entanglement and is given by
\begin{align}
N_A=\sum_{\lambda_n<0} |\lambda_n|, 
  \end{align}
where $\rho^{T_A}=\sum_n\lambda_n\ket{n}\bra{n}$ is the spectral decomposition of the partially transposed state with respect to  subsystem $A$. 
The total amount of correlations (classical and quantum) shared between the bipartitions $A$ and $B$ can be quantified using on the mutual information
\begin{equation}
  I(A:B)=S_{A}+S_B-S_{A\cup B},
\end{equation}
where $S_{A\cup B}$ is the von Neumann entropy of the state of the whole compound. 
In addition, we shall compute the coherence of the system state using the $L_1$ norm~\cite{Baumgratz:14}
\begin{align}
  C(\rho)=\sum_{i\neq j} |\rho_{i,j}|,
  \end{align}
where $\rho_{i,j}$ are the density matrix entries in the clock-state basis. Finally, we will use the quantifier of multipartite quantum correlations  provided by the so-called global quantum discord~\cite{Campbell:13,DeChiara:18}
\begin{align}\label{eq:G}
  \mathcal{G}(\rho)=\min_{\Pi^k}\left\{S(\rho||\Pi(\rho))-\sum_{i=1}^M S(\rho_i||\Pi_i(\rho_i)) \right\},
  \end{align}
where $\rho_i$ denotes the reduced state of the $i^\text{th}$ rotor, $\Pi(\rho)=\sum_j \Pi^j \rho \Pi^j$ is a projector operator acting on the global state, and  $\Pi_i(\rho_i)$ the corresponding projector acting on the single-rotor states. Following Ref.~\cite{Campbell:13}, we choose $\Pi^j=\mathcal{R}\ket{j} \bra{j} \mathcal{R}^\dagger$ 
with $\mathcal{R}=\otimes^M_{i=1} R_i({\bf \theta}_i)$ a collection of single-particle rotation operators, while the operator acting on the $i^\text{th}$ rotor reads
\begin{equation}
 R_i(\boldsymbol{\theta}_i)=e^{i \boldsymbol{\theta}_i \cdot \boldsymbol{\Lambda}},
\label{Ri:eq}
\end{equation} 
 where $ \boldsymbol{\theta}_i =(\theta_{i,1}\, ,\theta_{i,2}\dots \theta_{i,n_a})$ is a vector of $n_a$ angles, and $\boldsymbol{\Lambda}=(\Lambda_{1}\, ,\Lambda_{2}\dots \Lambda_{n_a})$ is a vector of generators of rotations for the single rotor. We have considered the $n_a=8$ Gell-Mann $3\times 3$ matrices as generators of rotations. The minimum in Eq.~(\ref{eq:G}) is obtained by varying the set of angles $\{\boldsymbol{\theta}_i\}$, $i=1,\dots M$, through an annealing algorithm.

\begin{figure}
\centering
\includegraphics[width=1\linewidth,angle=-0]{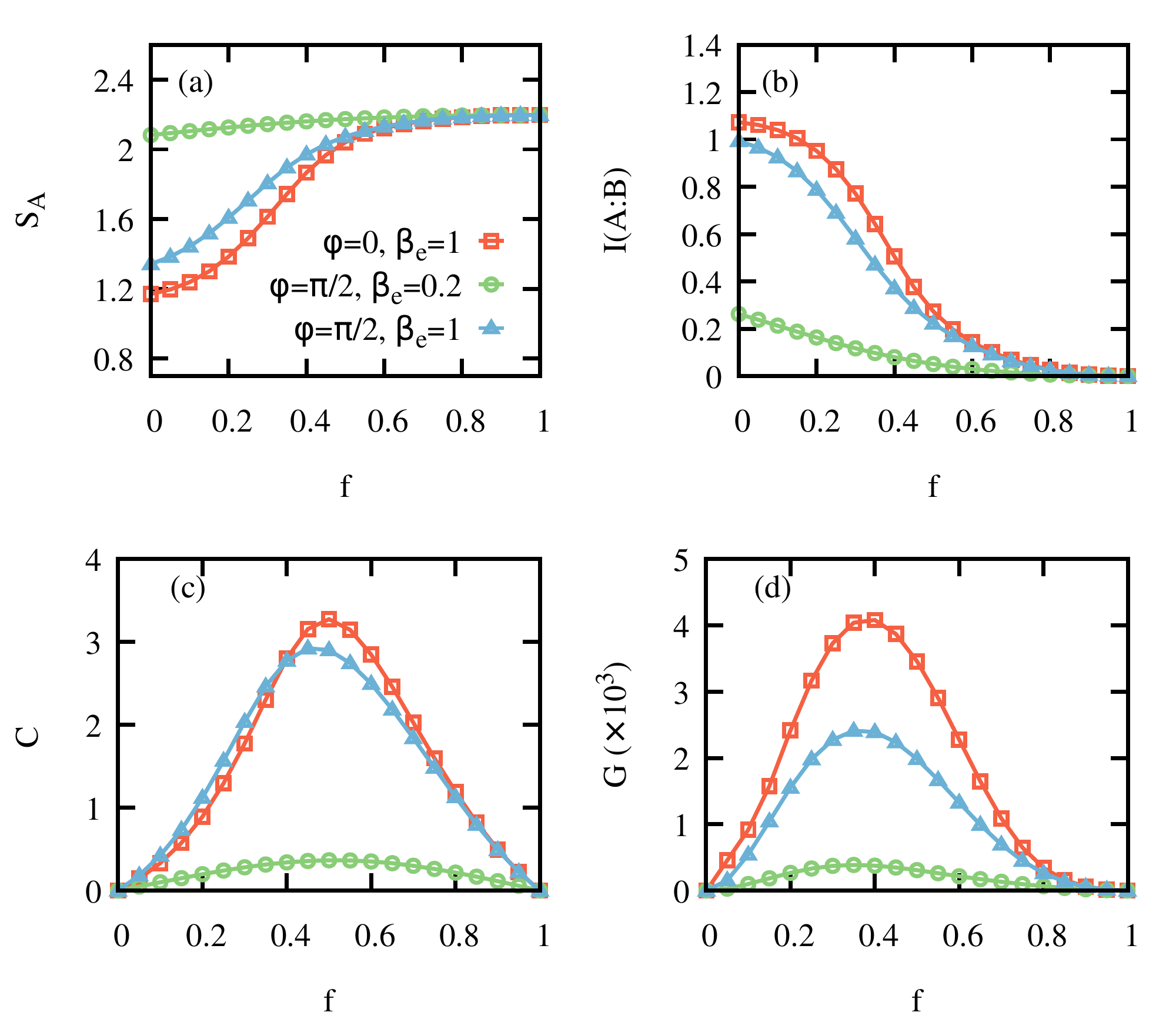}
\caption{\small{Information measures of the NESS for $M=4$ rotors as a function of the control parameter $f$. Panels (a), (b), (c) and (d) show the von Neumann entropy $S_A$, mutual information $I(A:B)$, coherence $C$ and global discord, respectively, with $A$ denoting the first half of the chain, namely, rotors $1$ and $2$. Different points (and colors) correspond to distinct values of the staggered chiral phase, such that $\varphi_j=(-1)^j\varphi$, and local temperature $\beta_{e}$, while $g=0.2$ and $\beta_o=1.1$.  See main text for further details.}}
\label{fig4}
\end{figure}

These instruments are all very informative of the quantum critical features of the ground-state QPT~\cite{Osborne:02,Vidal:03} in the CCM (cf. Appendix~\ref{app:a}). Here however we are mainly interested in the NESS properties: In such an open quantum system, critical features become blurred or disappear altogether. This might lead one to naively think that no connection could be established. Yet, the interplay between temperature gradient between sub-lattices, currents, and correlations reveal a rich phenomenology. Fig.~\ref{fig4} shows the behavior of these quantities for different parameters. Contrary to the CCM ground-state, these quantities show a smooth dependence on $f$, which suggests that it is not in partition-dependent quantities that a behavior reminiscent of a critical one should be sought. However, it is interesting to observe that both $S_A$ and $I(A:B)$ have an inflexion point in the region where we expect the critical value of $f$ to occur, which indicates a qualitative change in trend taking place around $f\simeq0.46$. On the other hand, the global quantum discord shows the quantumness acquired by the NESS away from $f=0,1$, which correlates with the amount of coherence $C$. However, while the coherence for the chiral model becomes maximum at $f\simeq 0.46$, the global discord peaks at a slightly  smaller value of $f$.
This should be compared with figures \ref{fig2} (a) and (b), and \ref{fig3}-(a) and (b), showing a similar behaviour for the tunnelling and the thermal current. This suggests that the amount of quantumness, as measured by the coherence and the global discord, is an essential ingredient for the out-of-equilibrium CCM to work as a thermal machine, thus converting thermal currents into mechanical currents.

Note however, that although $\rho_{\rm SS}$ contains coherence for $f\neq 0,1$, its maximum value is significantly smaller than in the ground-state where $C\propto N_s^M$ for $f>f_c$ [cf. Fig.~\ref{fig_appA2}(d)]. Similar behavior is observed for other choices of $\varphi$, also for $\varphi=k\pi/N_s$ with $k\in\mathbb{Z}$. In addition, all these quantities inherit the periodicity $2\pi/N_s$ in the phase $\varphi$. 
Finally, we stress that $N_{A}=0 \ \forall f,\varphi$, in contrast to the ground-state negativity (cf. App.~\ref{app:a}).

\begin{figure}
\centering
\includegraphics[width=1\linewidth,angle=-0]{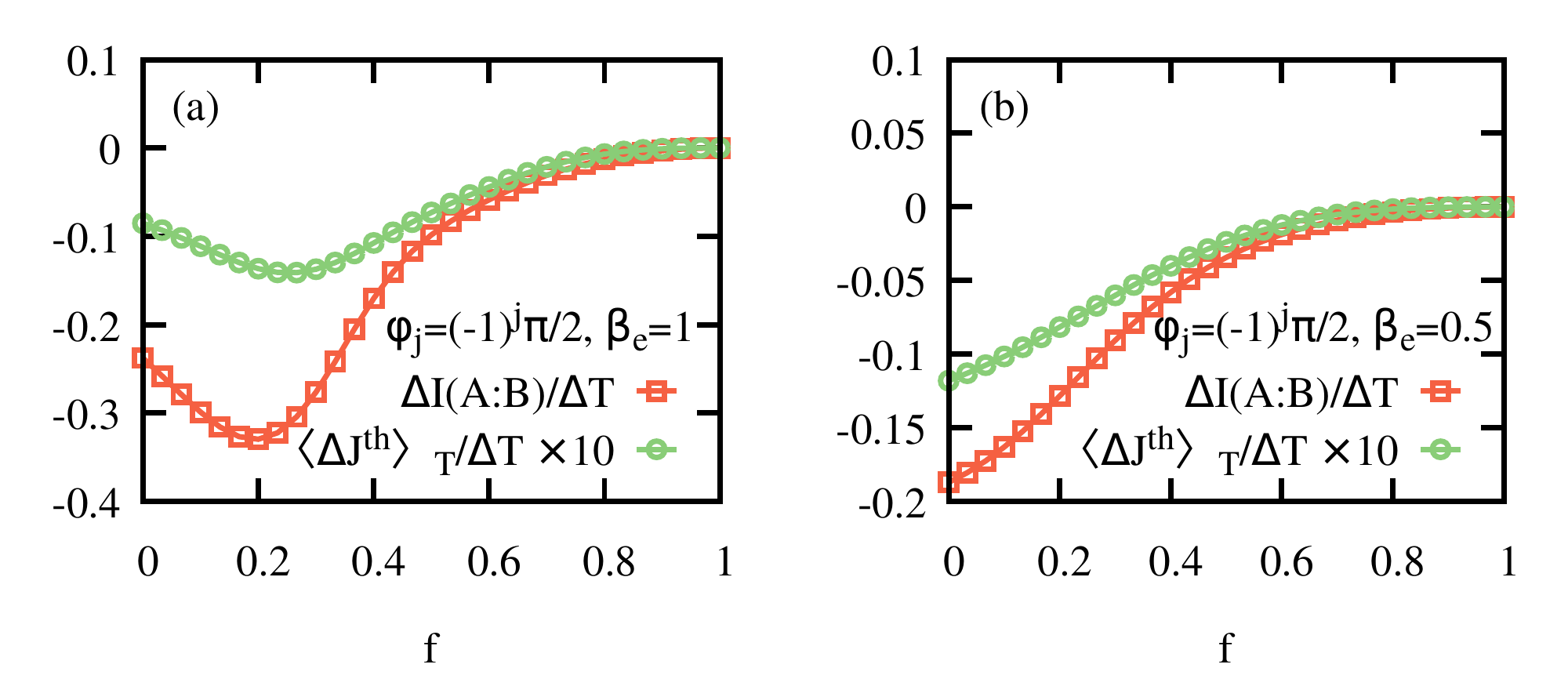}
\caption{\small{Mutual information and current susceptibilities, $\Delta I(A:B)/\Delta T$, $\langle \Delta J^{\rm th}\rangle_T/\Delta T$ as a function of the control parameter for $M=4$ rotors and $\varphi_j=(-1)^j\pi/2$. Panel (a) and (b) corresponds to $\beta_e=1$ and $\beta_e=0.5$, respectively, such that $\beta_o=1/(T_e+\Delta T)$ with $\Delta T=10^{-3}$, and $g=0.2$. The current susceptibility is multiplied by a factor $10$ for a better representation. Both quantities feature a qualitatively similar behavior.}}
\label{fig5}
\end{figure}

The observed behavior of the mutual information correlates with that of the total current $\langle J^{\rm th}+J^{\rm tun}\rangle_T\approx \langle J^{\rm th}\rangle_T$ (cf. Fig.~\ref{fig3}(b) and Fig.~\ref{fig4}(b)). Building on this observation, we investigate the thermal susceptibility of the total current $\langle \Delta J^{\rm th}\rangle_T/\Delta T$ with that of the mutual information, $\Delta I(A:B)/\Delta T$ when $\Delta T\rightarrow 0$, where $\langle \Delta J^{\rm th}\rangle_T\equiv \langle J^{\rm th}(\Delta T)\rangle_T-\langle J^{\rm th}(\Delta T=0)\rangle_T$ denotes the increment in the total current between $0<|\Delta T|\ll 1$ and $\Delta T=0$ (equal temperatures for both sub-lattices), and equivalently for the mutual information. Note that since $\langle J^{\rm th}(\Delta T=0)\rangle_T=0$, it follows $\langle \Delta J^{\rm th}\rangle_T/\Delta T=\langle  J^{\rm th}(\Delta T)\rangle_T/\Delta T$. In order to illustrate this susceptibility, we fix $\beta_e=1/T_e$ and change $\beta_o=1/(T_e+\Delta T)$ for $|\Delta T|\ll 1$. In Fig.~\ref{fig5} we show two examples of the mutual information and current susceptibility for $\Delta T=0.001$ and different $T_e$, for a fixed $\varphi_j=(-1)^j\pi/2$. Both susceptibilities feature a qualitative similar behavior, as well as for other choices of the parameters. We stress however that for different choices of $\varphi$ (or $\beta_o$) one may revert their relative sign. In addition, one should note that the total current vanishes for $\varphi=k\pi/N_s$ with $k\in\mathbb{Z}$, while the mutual information does not.

\section{Conclusions}\label{conc}
We have addressed the link between the emergence of NESS currents in a chiral few-body interacting-clock model and critical features of the corresponding model at the thermodynamic limit: The response of the system, in terms of currents, is maximum at the working point where a QPT is predicted to occur. This is also well captured by the behavior of genuinely multipartite information theoretic quantities, such as global quantum discord, and provides strong numerical evidences of the possible role that collective quantum phenomena play in the non-equilibrium response of this interesting interacting model. Such link will be explored further in future works through the investigation of possible effects in work-extraction games aimed at achieving ergotropic performance from the thermal-to-mechanical current-conversion process that we have addressed here. 

Furthermore, the investigation of the dynamical properties of the ground state of the model \eqref{eq:Hccm} and its variations is an interesting open question. In particular we find that tunnelling currents can arise in a rotated version of \eqref{eq:Hccm} with a finite number of rotors. Whether such currents persist in  the thermodynamic limit and exhibit a critical behaviour are questions worthy of future studies.

\begin{acknowledgments}
AI gratefully acknowledges the financial support of The Faculty of Science and Technology at Aarhus University through a Sabbatical scholarship and the hospitality of the Quantum Technology group, the Centre for Theoretical Atomic, Molecular and Optical Physics and the School of Mathematics and Physics, during his stay at Queen's University Belfast. AB acknowledges the hospitality of the Institute for Theoretical Physics and the ''Nonequilibrium quantum dynamics'' group at Universit{\"a}t Stuttgart, where part of this work was carried out. RP and MP acknowledge the support by the SFI-DfE Investigator Programme (grant 15/IA/2864), the Eropean Union's Horizon 2020 FET-Open project SuperQuLAN (899354) and TEQ (766900). MP acknowledges support by the Leverhulme Trust Research Project Grant UltraQuTe (grant RGP-2018-266) and the Royal Society Wolfson Fellowship (RSWF/R3/183013).  AB also acknowledges support from H2020 through the MSCA IF pERFEcTO (Grant Agreement nr. 795782) and from the DeutscheForschungsgemeinschaft (DFG, German Research Founda-tion) project number BR 5221/4-1. 
\end{acknowledgments}

\appendix

\begin{figure}
\centering
\includegraphics[width=1\linewidth,angle=-0]{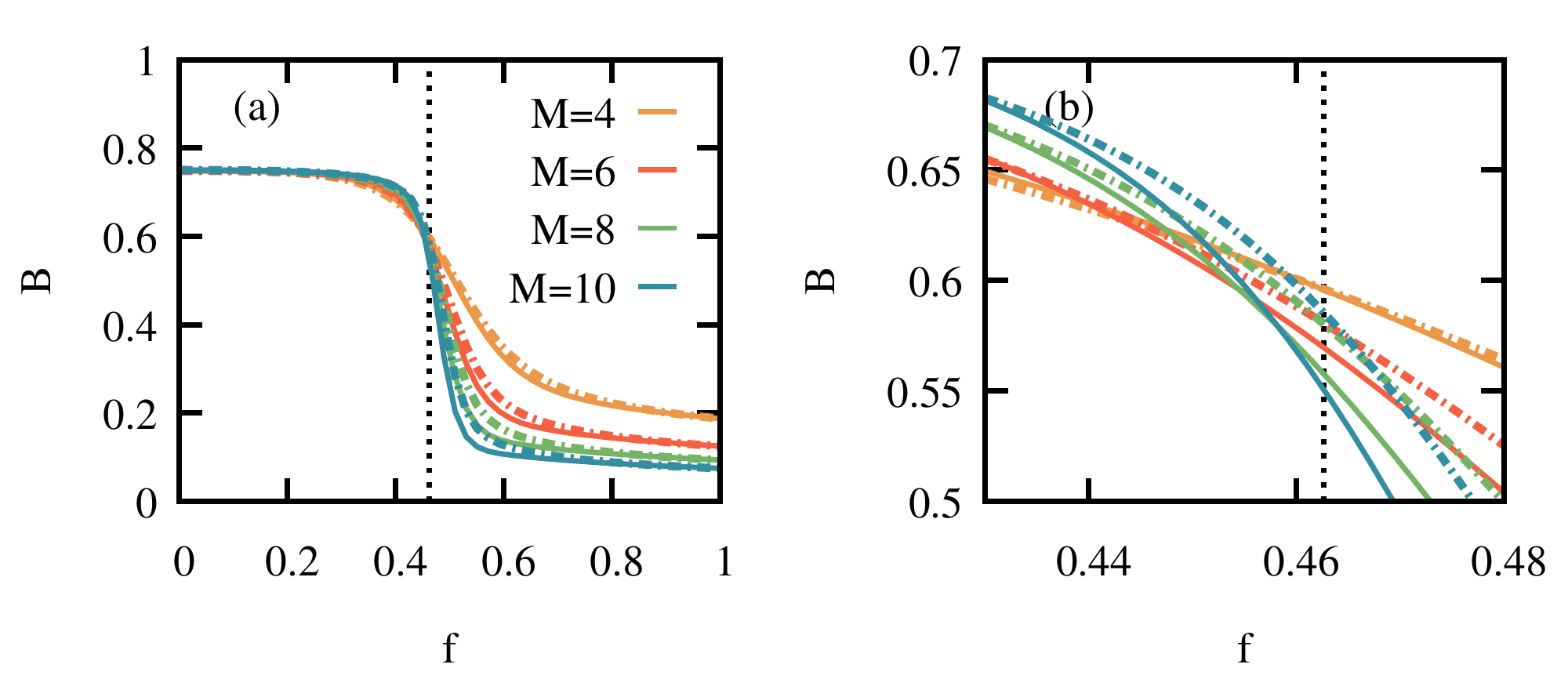}
\caption{\small{(a) Binder cumulant $B$ for the ground state of the CCM with $N_s=3$ for different system sizes (from $M=4$ to $10$ rotors) as function of $f$ and with $\varphi_j=\pi/8$ (solid lines) and staggered chiral phase $\varphi_j=(-1)^j\pi/8$ (dashed lines). The vertical dotted line indicates the critical value $f_c$ where the QPT takes place, reported in Ref.~\cite{Samajdar:18}. Panel (b) shows a zoom close to the region where Binder cumulants intersect (close to $f_c$). For homogeneous chiral phase, the the crossing approaches the reported value $f_c$. The intersections in $B$ for staggered chiral phases suggests that location of $f_c$ is shifted to a slightly larger value. }}
\label{fig_appA}
\end{figure}

\begin{figure*}
\centering
\includegraphics[width=1\linewidth,angle=-0]{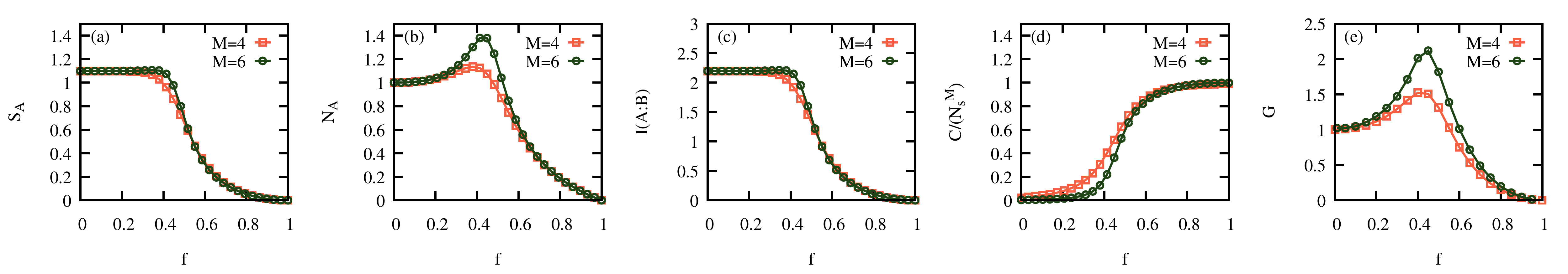}
\caption{\small{Quantum information measures of the ground state of the CCM with periodic boundary conditions and staggered chiral phase $\varphi_j=(-1)^j\pi/2$ for $M=4$ and $6$ rotors, which unveil the QPT taking place in the system. From left to right, von Neumann entropy $S_{A}$, negativity $N_A$,  mutual information $I(A:B)$, coherence $C$ (rescaled over the total Hilbert space dimension $N_s^M$), and global discord $\mathcal{G}$, respectively. The system is split in half, so the partition $A$ includes the first $M/2$ rotors, namely, rotors $1$ and $2$ for $M=4$ and $1$, $2$ $3$ for $M=6$. }}
\label{fig_appA2}
\end{figure*}

\begin{figure}
\centering
\includegraphics[width=1\linewidth,angle=-0]{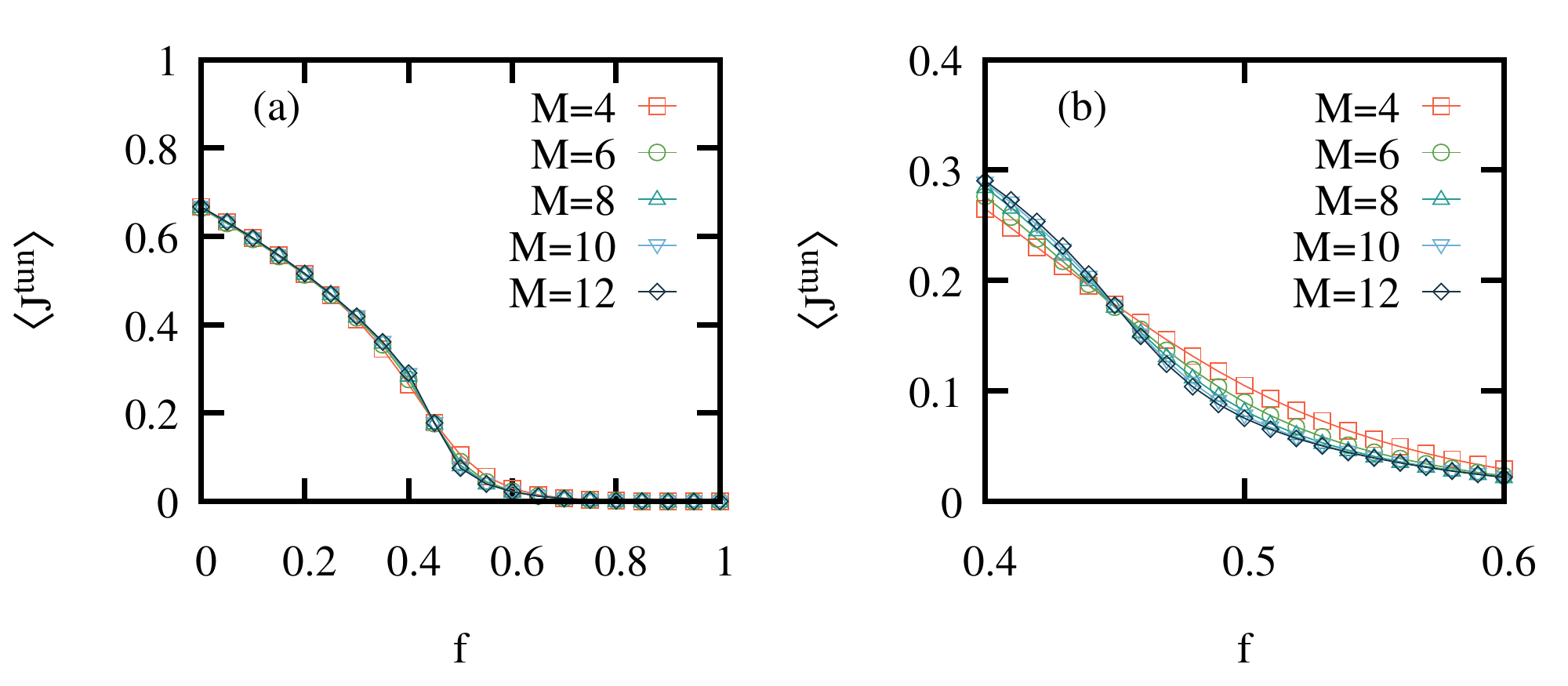}
\caption{\small{(a) Tunneling current for the odd rotors $\langle J^{\rm tun}\rangle$ in the ground state of the CCM $\tilde{H}_{\rm ccm}$, given in Eq.~\eqref{eq:Hsachdev}, with $\varphi_j=(-1)^j\pi/2$ and as a function of $f$. The tunneling current for even rotors is reversed in sign, i.e., $-\langle J^{\rm tun}\rangle$. Panel (b) shows a zoom closer to the transition point to signal the sharper behavior of $\langle J^{\rm tun}\rangle$ as $M$ increases.}}
\label{fig_appB}
\end{figure}

\section{Critical ground state features in the chiral clock model}\label{app:a}

As already noted, the CCM exhibits a $\mathbb{Z}_{N_s}$ symmetry. In order to exploit this symmetry,  it is handy to remap the Hamiltonian as $\sigma\rightarrow \mu$ and $\mu\rightarrow \sigma$ (as in Ref.~\cite{Samajdar:18}), so that the operator
\begin{align}
  \mathcal{U}=\prod_{j=1}^M \mu_j^\dagger,
\end{align}
allows us to split the Hilbert space in $N_s$ sub-spaces.  The ground state is contained in the subspace with eigenvalue $1$. In the case of $N_s=3$, the operator reads as $\mathcal{U}=\Pi_0 \Pi_0^\dagger +\omega \Pi_1 \Pi_1^\dagger + \omega^2\Pi_2 \Pi_2^\dagger$ where $\Pi_n$ denotes the projector on the corresponding subspace. One can use this symmetry to reduce the dimension of the Hilbert space. In particular, 
\begin{align}\label{eq:z3par}
H_{\rm ccm}^0= \Pi_0 H_{\rm ccm} \Pi_0^\dagger.
\end{align}
contains the ground state $\forall f$ with a well defined symmetry.  For $f<f_c$, the ground state becomes $N_s$-fold degenerate. 
In~\cite{Samajdar:18} the ground-state energy critical exponents of the $H_{\rm ccm}$ were investigated. For completeness, here we just provide a brief summary of the critical features of such a model. In particular, note that the ground-state order parameter $m=\frac{1}{M}\sum_{j=1}^M (\mu_j+\mu_j^\dagger)$ within the $\Pi_0$ subspace is given by
\begin{align}
\bra{\varphi(f)}m\ket{\varphi(f)}\equiv 0 \quad \forall f,
\end{align}
where $\ket{\varphi(f)}$ denotes the ground state of $H_{\rm ccm}^0$. As customary in symmetry-breaking phase transitions, one needs to resort to $m^2$ and $m^4$, which clearly reveal the symmetry-broken phase for $f<f_c$ (and thus the QPT). Moreover,  the location of the QPT can be witnessed by looking at the energy gap $\Delta$ or Binder cumulant $B$~\cite{Binder:81}. The energy gap between the ground and first excited state closes at $f_c$ following the universal scaling law~\cite{Sachdev}
\begin{align}
\Delta \sim |f-f_c|^{z\nu}.
\end{align}
where $z\nu$ are critical exponents of the QPT. The Binder cumulant is defined as~\cite{Binder:81}
\begin{align}
B=\frac{1}{2}\left(3-\frac{\langle m^4 \rangle}{\langle m^2\rangle^2}\right)
\end{align}
where $\langle m^4 \rangle$ and $\langle m^2\rangle$ are evaluated over the ground state, i.e. $\langle \varphi(f)|m^2|\varphi(f)\rangle$ and $\langle \varphi(f)|m^4|\varphi(f)\rangle$. This quantity has been proven very useful to locate the critical point $f_c$ (see for example Refs.~\cite{Angelini:14,Puebla:19}). Applying finite-size scaling arguments, $B$ is expected to become size independent at $f_c$. Hence, the QPT takes place at the value of $f$ at which the Binder cumulant $B$ for different system sizes $M$ intersect, although finite-size corrections still yield small deviations to the size-independent intersections. In Fig.~\ref{fig_appA} we show the resulting Binder cumulant $B$ for the ground state of the CCM for $N_s=3$ for the case of a staggered and homogeneous chiral phase $\varphi_j=(-1)^j\pi/8$ and $\varphi_j=\pi/8$, respectively. The location of the QPT, i.e. $f_c$, for the homogeneous chiral phase is consistent with the reported value in Ref.~\cite{Samajdar:18}, $f_c=0.46267$ which is indicated by a dotted vertical line, while $f_c$ appears to be shifted to a slightly larger value for a staggered chiral phase. The signatures of the QPT are already evident even for the considered system sizes $M\lesssim 10$. 

In addition, in Fig.~\ref{fig_appA2} we show the quantum information measures on the CCM ground-state as a function of the control parameter $f$, namely, von Neumann entropy $S_A$, negativity $N_A$, coherence $C$, mutual information $I(A:B)$ and global quantum discord $\mathcal{G}$. The system is split in half, so that $A$ refers to the first two rotors for $M=4$. All the quantities indicate a QPT taking place at $f\approx 0.46$~\cite{Samajdar:18}. Compare these ground-state results with those discussed in the main text for the NESS. 
It is worth noting that the non-chiral model $\varphi_j=0$ has a critical field $f_c=1/2$ \cite{Samajdar:18}, thus the chirality lowers the value of the field required to achieve the disordered phase, as one would expect.

\section{Tunneling current in a rotated CCM model}\label{app:c}
As commented in the main text, while the ground-state properties of the CCM model remain unaltered upon the rotation $\sigma\rightarrow \mu$ and $\mu\rightarrow \sigma$, the tunneling current becomes remarkably different. Note  that the definition of the tunneling current $J^{\rm tun}(j\rightarrow j')$ given in Eq.~\eqref{eq:Jtun} is independent of the specific choice of the Hamiltonian. In particular, for 
\begin{align}\label{eq:Hsachdev}
    \tilde{H}_{\rm ccm}=-f\sum_{j=1}^M(\mu_j+\mu_j^+)-(1-f)\sum_{j=1}^M\left(\sigma_j\sigma_{j+1}^\dagger e^{i\varphi_j}+{\rm H.c.}\right),
\end{align}
with staggered chiral phases $\varphi_j=(-1)^j\varphi$, the tunneling current in its ground state is non zero. Moreover, the behavior of $J^{\rm tun}$ resembles that of a critical quantity across a phase transition. This is plotted in Fig.~\ref{fig_appB} for $M$ from $4$ to $12$ rotors, which indicate a sharp transition around the QPT.


%

\end{document}